# Linking emergent phenomena and broken symmetries through one-dimensional objects and their dot/cross products

Sang-Wook Cheong[1], Fei-Ting Huang[1] and Minhyong Kim[2]

[1] Rutgers Center for Emergent Materials and Department of Physics and Astronomy, Rutgers University, 136 Frelinghuysen Rd, Piscataway, New Jersey, USA
[2] International Centre for Mathematical Sciences, The Bayes Centre, University of Edinburgh, 47 Potterrow Edinburgh EH8 9BT, UK and Korea Institute for Advanced Study, 85 Hoegiro, Dongdaemun-gu, Seoul, Korea
E-mail: sangc@physics.rutgers.edu

**Abstract**

The symmetry of the whole experimental setups, including specific sample environments and measurables, can be compared with that of specimens for observable physical phenomena. We, first, focus on one-dimensional (1D) experimental setups, independent from any spatial rotation around one direction, and show that eight kinds of 1D objects (four; vector-like, the other four; director-like), defined in terms of symmetry, and their dot and cross products are an effective way for the symmetry consideration. The dot products form a $Z_2 \times Z_2 \times Z_2$ group with Abelian additive operation, and the cross products form a $Z_2 \times Z_2$ group with Abelian additive operation or $Q_8$, a non-abelian group of order eight, depending on their signs. Those 1D objects are associated with characteristic physical phenomena. When a 3D specimen has Symmetry Operational Similarity (SOS) with (identical or lower, but not higher, symmetries than) an 1D object with a particular phenomenon, the 3D specimen can exhibit the phenomenon. This SOS approach can be a transformative and unconventional avenue for symmetry-guided materials designs and discoveries.

## 1. Introduction

Symmetry often governs the rules of this universe. In condensed matter physics, we often take into account symmetry in (magnetic) point/space groups [1-2] and their tensorial response functions, Hamiltonian, and also through the Ginzburg–Landau approach [3]. Order parameters, measuring the degree of order across phase transitions that accompany symmetry breaking, are often vectorlike entities. Examples include magnetization, polarization, ferro-rotational axial vector, and magnetic toroidal moment [4-7]. All of these have broken 2-fold rotation around any axis perpendicular to the vectorial direction. However, order parameters for ferroelastic or crystallographic chiral (screw-like) transition are director-like, i.e. invariant under 2-fold rotation around any axis perpendicular to the directorial direction. Indeed, Hlinka [8] identified "eight" distinct types of vectorlike entities, each of which has characteristic properties under various symmetry operations. According to the Hlinka classification, there exist two additional kinds of vectorlike entities, in addition to magnetization, polarization, ferro-rotational axial vector, magnetic toroidal moment, ferroelastic strain and chirality.

For the practical utilization of the eight kinds of vectorlike entities for real 3D materials and observable physical phenomena, the concept of Symmetry Operational Similarity (SOS) was introduced and a large number of quasi-one-dimensional (quasi-1D) objects, exhibiting SOS with each type of vectorlike entities, have been identified [9-11]. It turns out that numerous physical phenomena can be systematically and succinctly understood in terms of the SOS concept and these quasi-1D objects. The primary reason for the power of this approach is that what we experimentally measure is often associated with (quasi-)1D objects such as magnetization, electric polarization, unidirectional current flow or unidirectional light propagation. SOS relationship between a quasi-1D object and one of eight vectorlike entities means that the quasi-1D object, often placed in the left-hand side, has identical or lower symmetries than the vectorlike entity, often placed in the right-hand side. In other words, in order to have an SOS relationship, the left-hand-side object cannot have higher symmetries than the right-hand-side vectorlike entity. It turns out that the exact meaning of quasi-1D objects have never been discussed in terms of symmetry. In this paper, we define "1D objects" rigorously in terms of symmetry, derive the full list of dot and cross products between 1D objects, and consider their group theoretical and algebraic relationships. Note that traditionally, 1D objects are defined in terms of interaction or connectivity, but we will define it in terms of symmetry, which will be crucial for our group theoretical analysis. We also discuss how 1D object classification and their dot and cross products can be utilized to explain and also predict new phenomena in 3D materials with broken symmetries.

As we will discuss extensively, our SOS approach with 1D objects is a succinct manner to link between emergent phenomena and broken symmetries of complex 3D materials. We may ask a variety of questions on emergent complex phenomena such as these: (1) Can we have "natural" optically activity in magnetic states? (2) What can strain gradient do on linear magnetoelectrics? (3) Can magnetic field gradient do anything interesting in crystalline solids? (4) What can couple with vortex light beams, so interesting phenomena can be observed with vortex light beams? (5) Can thermal gradient be coupled with magnetic states? (6) Can we observe Magneto-Optical Kerr Effect (MOKE) in antiferromagnetic states? (7) Can we observe off-diagonal voltage with current flow without magnetic field? All of these questions as well as many more can be readily answered with our SOS approach with 1D objects. Furthermore, the relevant broken symmetries can be identified through our approach, so the relevant magnetic point groups and materials can be also identified.

First, we use the general symmetry operation notations for three orthogonal $x$, $y$, and $z$ axes such as $\mathbf{R_x}$=2-fold rotation around the $x$ axis, $\mathbf{M_x}$=mirror reflection with mirror perpendicular to the $x$ axis, $\mathbf{I}$=space inversion, $\mathbf{T}$=time reversal, etc. Then, we have these general relationships: $\mathbf{R_x} \otimes \mathbf{R_y} = \mathbf{R_z}$, $\mathbf{M_z} \otimes \mathbf{R_y} = \mathbf{M_x}$, $\mathbf{M_x} \otimes \mathbf{R_z} = \mathbf{M_y}$, $\mathbf{M_x} \otimes \mathbf{R_y} = \mathbf{M_z}$, $\mathbf{M_y} \otimes \mathbf{M_z} = \mathbf{R_x}$, $\mathbf{M_x} \otimes \mathbf{M_z} = \mathbf{R_y}$, $\mathbf{M_x} \otimes \mathbf{M_y} = \mathbf{R_z}$, $\mathbf{M_x} \otimes \mathbf{R_x} = \mathbf{M_y} \otimes \mathbf{R_y} = \mathbf{M_z} \otimes \mathbf{R_z} = \mathbf{I}$, $\mathbf{M_x} = \mathbf{R_x} \otimes \mathbf{I}$, $\mathbf{M_y} = \mathbf{I} \otimes \mathbf{R_y}$,
and $\mathbf{M_z} = \mathbf{I} \otimes \mathbf{R_z}$ (all of these are commutative.).

With the above symmetry notations, we define 1D objects along the $z$ axis precisely in terms of the following three symmetry requirements: (We call this $z$ axis as the 1D direction of 1D objects. Our symmetry discussion will be most useful for crystalline solids with lattice, i.e. periodic structures and vectorlike entities, but these 1D objects are not limited to those.)

***Requirement (1)*** $\mathbf{R_z}$ is identity, i.e. unbroken. It may have additional rotational symmetry (such as $C_3$, $C_4$, $C_6$, or $C_\infty$) around the $z$ axis. In other words, it has, at least, $C_2$ symmetry around the $z$ axis.

***Requirement (2)*** $\mathbf{R_x}=\mathbf{R_y}$, $\mathbf{M_x}=\mathbf{M_y}$, and their symmetry properties should not depend on the orientation of the x or y axis. In addition, it does not have any rotational symmetry (such as $C_3$ or $C_4$), around any axis, different from the $z$ axis, other than possible $C_2$ around the $x$ or $y$ axis. In other words, it has, at most, $C_2$ symmetry around the $x$ or $y$ axis.

***Requirement (3)*** Any symmetry breaking is only in one manner, so the combination of two broken symmetry operations always results in an unbroken state.

Note that in all of our symmetry considerations, translation is ignored, or freely allowed, and in the symmetry consideration in Requirements (2) and (3), any rotation around the $z$ axis is also ignored, or freely allowed.

With the above definition for 1D objects, $\mathbf{R_x}$ and $\mathbf{I}$ are sufficient to tell all symmetry properties of 1D objects except time reversal, since $\mathbf{R_z}=1$, $\mathbf{R_y}=\mathbf{R_x}$, $\mathbf{M_x}=\mathbf{M_y}=\mathbf{I}\otimes\mathbf{R_x}$, and $\mathbf{M_z}=\mathbf{I}\otimes\mathbf{R_z}=\mathbf{I}$. Therefore, we can have only these choices for 1D objects: broken or unbroken $\mathbf{R_x}$, $\mathbf{I}$ and $\mathbf{T}$, so $2^3=8$ possibilities. This is why there are eight kinds of 1D objects [10]. Note that all 1D objects along the $z$ axis do have $\mathbf{R_z}$, and we do not consider any high-even-order rotational symmetry (such as $C_4$, $C_6$, or $C_\infty$) around the $z$ axis.

Here, we list a number of examples of 1D objects and also non-1D objects, many of which are counter-intuitive:
(1) Arrow shaped objects and double arrow shaped objects are 1D objects.
(2) A spring along the z axis is an 1D object along any of the $x$, $y$, and $z$ axes.
(3) Vectors like ***E*** (electric field), ***H*** (magnetic field), and ***v*** (velocity) are 1D objects. However, two coexisting, for example, ***E*** and ***H*** along a direction without mutual coupling is not an 1D object, but are two 1D objects.
(4) Square bars without any internal structure (in Fig. 1(d) with square cross section) are 1D objects only along the $z$ axis. However, rectangular bars without any internal structure (in Fig. 1(d) with rectangular cross section) are 1D objects along any of the $x$, $y$, and $z$ axes. But triangular bars without any internal structure (in Fig. 1(e) with equilateral-triangular cross section) are not 1D objects due to broken $\mathbf{R_z}$.
(5) An antiferromagnetic chain along the $z$ axis such as Fig. 1(f) is an 1D object along the $z$ axis; however, an antiferromagnetic chain along the $z$ axis such as Fig. 1(g) is an 1D object along any of the $x$, $y$, and $z$ axis. Interestingly, a ferromagnetic chain with magnetization perpendicular to the chain direction is an 1D object along the ferromagnetic magnetization direction, not along the chain direction.
(6) Figure 1(i) ferro-rotation, (j) electric quadrupole, (k) magnetic toroidal moment, and (l) magnetic quadrupole are 1D objects.
(7) Monolayer h-BN ($P\bar{6}2m$) is not an 1D object due to broken $\mathbf{R}_z$, but bulk boron nitride with AB stacking ($P6_3/mmc$) is an 1D object along the stacking direction [12] as shown in Fig. 1(m) and (n). Similarly, H-MoS$_2$ monolayer ($P\bar{6}m2$) is not an 1D object, but bulk H-MoS$_2$ ($P6_3/mmc$) is an 1D object [13]. Black phosphorus ($Cmca$) and black phosphorene ($Pmna$) are 1D objects, but blue phosphorene is not an 1D object ($P\bar{3}m1$) [14]. However, both graphene ($P6/mmm$) and bulk graphite ($P6_3/mmc$) are 1D objects [15], and both 1T-TaS$_2$ monolayer ($P\bar{3}m1$) and bulk 1T-TaS$_2$ ($P\bar{3}m1$) are not 1D objects [13].
(8) When red arrows in Fig. 1(o) are considered as lattice distortions or local electric polarization, then the lattice configuration has $\bar{4}=C_4\otimes\mathbf{I}$ symmetry. It has broken $\mathbf{R}_x=\mathbf{R}_y$ and $\mathbf{R}_z$ is unbroken, so it is an 1D object. When blue arrows in Fig. 1(p) are considered as spins, then the spin configuration has unbroken $\bar{4}'=\bar{4}\otimes\mathbf{T}=C_4\otimes\mathbf{I}\otimes\mathbf{T}$. It has unbroken $\mathbf{R}_z$, but has broken all of $\mathbf{R}_x=\mathbf{R}_y$, so it is an 1D object.
(9) Point groups of 2, 4 and 6 do have only $C_2$, $C_4$, and $C_6$ symmetries around the $z$ axis, respectively. In other words, all of $\mathbf{I}$, $\mathbf{R}_x$ and $\mathbf{M}_x=\mathbf{I}\otimes\mathbf{R}_x$ are broken, so they are not 1D objects. Point group of 23 has unbroken $\mathbf{R}_x$, $\mathbf{R}_y$, and $\mathbf{R}_z$, and also unbroken $C_3$ rotation around any of [111]-type directions. Thus, it is not an 1D object. Point groups of 222, 422, and 622 do have $C_2$ symmetry around any of the $x$, $y$, and $z$ axis, and has broken $\mathbf{M}_x=\mathbf{I}\otimes\mathbf{R}_x=\mathbf{I}$. Thus, they are 1D objects.

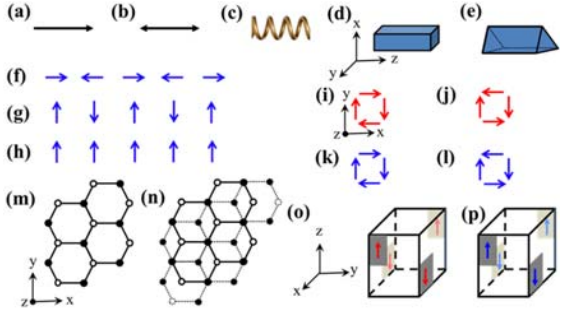

**Figure 1.** 1D objects vs. non-1D objects. (a) arrow shape, (b) double arrow shape, (c) spring, (d) rectangular bar without any internal structure, (e) triangular bar without any internal structure, (f) & (g) antiferromagnetic chain, (h) ferromagnetic chain with magnetization perpendicular to the chain direction, (i) ferro-rotation, (j) electric quadrupole, (k) magnetic toroidal moment, (l) magnetic quadrupole, (m) monolayer h-BN, and (n) h-BN with AB stacking along the $z$ axis, (o) lattice with local lattice distortions with $\bar{4}$ symmetry, (p) spin structure with $\bar{4}\otimes\mathbf{T}$ symmetry. Red and blue arrows are local electric polarizations and spins, respectively. Many of these cases are counter-intuitive for being 1D objects along a particular direction – see the text for the correct identifications.

We emphasize that in the above consideration of 1D objects along the $z$ axis, we ignore or freely allow the rotation around the $z$ axis, especially for Requirements (2) and (3), so the real symmetry of, for example, 3D point groups can be lower, but not higher, than what is described in the above.

## 2. Results
### 2.1 Symmetry properties of 1D objects

We now discuss the symmetry operational properties of 1D objects using these symmetry operation notations: **R**=2-fold rotation with the rotation axis perpendicular to the 1D direction, **R**=2-fold rotation with the rotation axis along the 1D direction, **I**=space inversion (parity), **M**=mirror reflection with the mirror perpendicular to the 1D direction, **M**=mirror reflection with the mirror plane containing the 1D direction, and **T**=time reversal. Assume the 1D object lies along the $z$ axis.
(1) **R**: 2-fold rotation around the $x$ or $y$ axis, **R**: $(x,y,z) \rightarrow (x,-y,-z)$ or $(x,y,z) \rightarrow (-x,y,-z)$.
(2) **R**: 2-fold rotation around the $z$ axis, R = $(x, y, z) \rightarrow (-x, -y, z)$.
(3) **M**: Mirror reflection across the $xy$ plane, **M**: $(x,y,z) \rightarrow (x,y,-z)$.
(4) **M**: Mirror reflection across the $xz$ or $yz$ plane, **M**: $(x,y,z) \rightarrow (x,-y,z)$ or $(x,y,z) \rightarrow (-x,y,z)$.
(5) **I**: Space inversion, **I**: $(x, y, z) \rightarrow (-x,-y,-z)$.
Thus, we have relations like $\mathbf{I}=\underline{\mathbf{M}}\otimes\mathbf{R}=\mathbf{M}\otimes\underline{\mathbf{R}}$.
In Ref. [10], the following broken symmetries are listed for the Hlinka's eight 1D vectorlike entities:
𝒟: (Director) no broken symmetry
𝒶: (Axiality due to ferro-rotation) broken {**R**,**M**}
𝒞: (Chirality) broken {**I**,**M**,**M**}
𝒫: (Polarization) broken {**R**,**I**,**M**}
𝒟′: (Director′) broken {**T**}
𝒶′: (Axiality′) broken {**R**,**M**,**T**}
𝒞′: (Chirality′) broken {**I**,**M**,**M**,**T**}
𝒫′: (Polarization′) broken {**R**,**I**,**M**,**T**}
Note that [′] means broken **T**, and 𝒶, 𝒞, 𝒫, 𝒶′, 𝒞′, and 𝒫′ are, for example, associated with phenomena such as linear gyration, natural optical activity, non-zero electric polarization (or ferroelectricity), non-zero net magnetic moment (or ferromagnetism), linear magnetoelectricity, and directional non-reciprocity, respectively [10-11].

The symmetry operations are not listed above means that they are unbroken. However, it turns out that some of these broken symmetries are dependent to each other. For example, $\underline{M}=R\otimes I$, so when **I** is unbroken for $\mathcal{A}$, then $R=\underline{M}$. As we have discussed earlier, we know that the symmetry operations of **R**, **I**, and **T** are sufficient to tell all symmetry properties of 1D objects when high-even-order rotations along the 1D direction are ignored or freely allowed. If we will use 0 for unbroken symmetry (blue) and 1 for broken symmetry (red), then we have this list:

$\mathcal{D}$: (**R**,**I**,**T**)=(0,0,0)
$\mathcal{A}$: (**R**,**I**,**T**)=(1,0,0)
$\mathcal{C}$: (**R**,**I**,**T**)=(0,1,0)
$\mathcal{P}$: (**R**,**I**,**T**)=(1,1,0)
$\mathcal{D}'$: (**R**,**I**,**T**)=(0,0,1)
$\mathcal{A}'$: (**R**,**I**,**T**)=(1,0,1)
$\mathcal{C}'$: (**R**,**I**,**T**)=(0,1,1)
$\mathcal{P}'$: (**R**,**I**,**T**)=(1,1,1)

The 1D objects in Fig. 1 can be classified like these:
(1) Arrows: $\mathcal{P}$, double arrows: $\mathcal{D}$.
(2) spring: $\mathcal{C}$.
(3) *E*: $\mathcal{P}$, *H*: $\mathcal{A}'$, *v*: $\mathcal{P}'$.
(4) square or rectangular bars without any internal structure (in Fig. 1(d)): $\mathcal{D}$.
(5) An antiferromagnetic chain: $\mathcal{D}$, a ferromagnetic chain with magnetization perpendicular to the chain direction: $\mathcal{A}'$ along the ferromagnetic magnetization direction.
(6) Ferro-rotation in Fig. 1(i) has broken **R**, so is $\mathcal{A}$, and toroidal moment in Fig. 1(k) has broken **R**, **I**, and **T**, so is $\mathcal{P}'$. Fig. 1(j) and Fig. 1(l): $\mathcal{D}$.
(7) Boron nitride with AB stacking, graphene, black phosphorus, black phosphorene, bulk graphite, and bulk H-MoS$_2$: $\mathcal{D}$.
(8) The lattice configuration with $\bar{4}=C_4\otimes I$ symmetry in Fig. 1(o) has unbroken I when $C_4$ is freely allowed, so is $\mathcal{A}$. The spin configuration with $\bar{4}'=\bar{4}\otimes T=C_4\otimes I\otimes T$ symmetry in Fig. 1(p) has broken **I** and **T**, so is $\mathcal{P}'$.
(9) Crystalline point groups of 222, 422, and 622 have unbroken **R** and broken **I**, so they are $\mathcal{C}$. Note that all 32 crystalline point groups are classified as non-1D objects or 1D objects ($\mathcal{D}$, $\mathcal{A}$, $\mathcal{C}$, and $\mathcal{P}$) as shown in Table 1. This classification is based on SOS relationships. In other words, some of the 3D crystallographic point groups having SOS with a specific type of 1D objects can have lower symmetries than the specific type of 1D objects. For example, point group $\bar{4}$ has broken **I**, but $\mathcal{A}$ has unbroken **I**.

Emphasize that non-1D objects can often have SOS with an 1D object. For example, antiferromagnetic Cr$_2$O$_3$ with magnetic space group of $R\bar{3}'c'$ has unbroken $R_x$ and $R_y$, but broken $R_z$, so is not an 1D object along *z*. Now, $\mathcal{C}'$ has unbroken $\{R_x,R_y\}$ and broken $\{I,T\}$, so has broken "$\{R_x,R_y\}$ + unbroken $\{R_x,R_y\}\otimes$broken $\{I,T\}$" = broken $\{I,T,R_x\otimes I,R_y\otimes I,R_x\otimes T,R_y\otimes T\}$ = broken $\{I,T,M_x,M_y,R_x\otimes T,R_y\otimes T\}$. In fact, Cr$_2$O$_3$ has broken all of these symmetries, so does have SOS with $\mathcal{C}'$ along *z* – consequently, exhibit linear magnetoelectricity along *z* [16-17]. Similarly, Cr$_2$O$_3$ have SOS with $\mathcal{C}'$ along *x* and *y* as well, thus it can have all diagonal linear magnetoelectric tensor terms.

We also make one general note for SOS relationship: for example, $\mathcal{P}'$ with broken **T** does not have SOS with $\mathcal{P}$ with unbroken **T** since all 1D objects are distinct. Or, we may say that $\mathcal{P}'$ has unbroken $I\otimes T$, but $\mathcal{P}$ has broken $I\otimes T$, so $\mathcal{P}'$ does not have SOS with $\mathcal{P}$. Note that, for example, broken (**R**,**I**,**T**)=(1,0,1) for $\mathcal{A}'$ means that **I** is unbroken, **T** is broken and consequently $I\otimes T$ is broken for $\mathcal{A}'$. Also note that all eight types of 1D objects are distinct; for example, $\mathcal{A}'$ is not a subset of $\mathcal{D}'$ because **R** and $\underline{M}$ are unbroken in $\mathcal{D}'$, but broken in $\mathcal{A}'$. Another important aspect is: all of $\mathcal{D}$, $\mathcal{C}$, $\mathcal{D}'$, and $\mathcal{C}'$ have unbroken **R**, so they are director-like while $\mathcal{P}$, $\mathcal{A}$, $\mathcal{P}'$, and $\mathcal{A}'$ have broken **R**, so they are vector-like. From now on, we will call $\mathcal{D}$, $\mathcal{C}$, $\mathcal{D}'$, and $\mathcal{C}'$ as directors and $\mathcal{P}$, $\mathcal{A}$, $\mathcal{P}'$, and $\mathcal{A}'$ as vectors for the sake of simplicity (even though directors mean specifically only $\mathcal{D}$).

**Table 1**, Non-1D objects (gray) and 1D objects ($\mathcal{D}$, $\mathcal{A}$, $\mathcal{C}$, and $\mathcal{P}$) of crystalline point groups. Symbol daggers † indicate that space groups can have the 1D direction in any direction. Note that some of the 3D point groups having SOS with a specific type of 1D objects can have lower symmetries than the specific type of 1D objects. For example, $\bar{4}$, which belongs to $\mathcal{A}$, has broken **I**, but $\mathcal{A}$ has unbroken **I**. However, note that $\bar{4}$ is not $\mathcal{P}$ because $I\otimes C_4$ is unbroken in $\bar{4}$, but is broken in $\mathcal{P}$.

|  | non-1D objects | $\mathcal{D}$ | $\mathcal{A}$ | $\mathcal{C}$ | $\mathcal{P}$ |
|---|---|---|---|---|---|
| Cubic | 23, $m\bar{3}$, 432, $\bar{4}3m$, $m\bar{3}m$ |  |  |  |  |
| Hexagonal | 6, $\bar{6}$, $\bar{6}m2$ | 6/*mmm* | 6/*m* | 622 | 6*mm* |
| Trigonal | 3, $\bar{3}$, 32, 3*m*, $\bar{3}m$ |  |  |  |  |
| Tetragonal | 4 | 4/*mmm*, $\bar{4}2m$ | $\bar{4}$, 4/*m* | 422 | 4*mm* |
| Orthorhombic |  |  | *mmm*† |  | 222† | *mm*2 |
| Monoclinic |  | 2, *m* |  | 2/*m* |  |
| Triclinic |  | 1, $\bar{1}$ |  |  |  |

## 2.2 Dot products of 1D objects

Now, we can define dot products and cross products among these eight types of 1D objects: A dot product between two 1D objects along a spatial direction (**X**•**Y**=**Z**) means that two 1D objects (**X** and **Y**) coexist and are coupled to each other in such a way that they act like one (**Z**) of eight 1D objects along the same direction, and when a symmetry is broken in both 1D objects (**X** and **Y**), the 3rd 1D object (**Z**) has unbroken the symmetry. For example, we have $\mathcal{C}\bullet\mathcal{A}'$ = Broken {**I**,**M**} ⊗ Broken {**R**,**M**,**T**}= Broken {**R**,**I**,**T**} = $\mathcal{P}'$. Another example is $\mathcal{A}\bullet\mathcal{A}'$ = Broken {**R**,**M**} ⊗ Broken {**R**,**M**,**T**} = Broken {**T**} = $\mathcal{D}'$. The full list of dot products between two vectors/directors is shown in Table 2. Note that all dot products are commutative and associative, i.e. **X**•**Y**=**Y**•**X** and (**X**•**Y**) •**Z**=**X**•(**Y**•**Z**). For example, we have $\mathcal{C}\bullet\mathcal{P}'=\mathcal{P}'\bullet\mathcal{C}=\mathcal{A}'$ and $(\mathcal{C}\bullet\mathcal{P}')\bullet\mathcal{A}'=\mathcal{C}\bullet(\mathcal{P}'\bullet\mathcal{A}')=\mathcal{D}$. Also note that $\mathcal{D}$ acts like a unity.

**Table 2**, Dot products between two 1D objects

|  | $\mathcal{D}$ | $\mathcal{A}$ | $\mathcal{C}$ | $\mathcal{P}$ | $\mathcal{D}'$ | $\mathcal{A}'$ | $\mathcal{C}'$ | $\mathcal{P}'$ |
|---|---|---|---|---|---|---|---|---|
| $\mathcal{D}$ | $\mathcal{D}$ | $\mathcal{A}$ | $\mathcal{C}$ | $\mathcal{P}$ | $\mathcal{D}'$ | $\mathcal{A}'$ | $\mathcal{C}'$ | $\mathcal{P}'$ |
| $\mathcal{A}$ | $\mathcal{A}$ | $\mathcal{D}$ | $\mathcal{P}$ | $\mathcal{C}$ | $\mathcal{A}'$ | $\mathcal{D}'$ | $\mathcal{P}'$ | $\mathcal{C}'$ |
| $\mathcal{C}$ | $\mathcal{C}$ | $\mathcal{P}$ | $\mathcal{D}$ | $\mathcal{A}$ | $\mathcal{C}'$ | $\mathcal{P}'$ | $\mathcal{D}'$ | $\mathcal{A}'$ |
| $\mathcal{P}$ | $\mathcal{P}$ | $\mathcal{C}$ | $\mathcal{A}$ | $\mathcal{D}$ | $\mathcal{P}'$ | $\mathcal{C}'$ | $\mathcal{A}'$ | $\mathcal{D}'$ |
| $\mathcal{D}'$ | $\mathcal{D}'$ | $\mathcal{A}'$ | $\mathcal{C}'$ | $\mathcal{P}'$ | $\mathcal{D}$ | $\mathcal{A}$ | $\mathcal{C}$ | $\mathcal{P}$ |
| $\mathcal{A}'$ | $\mathcal{A}'$ | $\mathcal{D}'$ | $\mathcal{P}'$ | $\mathcal{C}'$ | $\mathcal{A}$ | $\mathcal{D}$ | $\mathcal{P}$ | $\mathcal{C}$ |
| $\mathcal{C}'$ | $\mathcal{C}'$ | $\mathcal{P}'$ | $\mathcal{D}'$ | $\mathcal{A}'$ | $\mathcal{C}$ | $\mathcal{P}$ | $\mathcal{D}$ | $\mathcal{A}$ |
| $\mathcal{P}'$ | $\mathcal{P}'$ | $\mathcal{C}'$ | $\mathcal{A}'$ | $\mathcal{D}'$ | $\mathcal{P}$ | $\mathcal{C}$ | $\mathcal{A}$ | $\mathcal{D}$ |

Emphasize that directors like $\mathcal{D}$, $\mathcal{C}$, $\mathcal{D}'$, $\mathcal{C}'$ are 1D object along any of the three orthogonal principal axes, unless there is a high-even-order rotational symmetry (such as $C_4$, $C_6$, or $C_\infty$) axis. In general, in dot products, directors like $\mathcal{D}$, $\mathcal{C}$, $\mathcal{D}'$, $\mathcal{C}'$ does not always have one well-defined 1D direction, and can orient in any directions. For example, in $\mathcal{C}\bullet\mathcal{A}'=\mathcal{P}'$, vectors like $\mathcal{A}'$ and $\mathcal{P}'$, have to be in the same direction, but $\mathcal{C}$ does not have to be in the same direction, and can, in fact, be in any direction, which is shown schematically in Fig. 2. We now have elegant algebraic relationships for dot products. For example, $\mathcal{C}\bullet\mathcal{P}'=\mathcal{P}'\bullet\mathcal{C}=\mathcal{A}'$ corresponds to (0,1,0)+(1,1,1)=(1,1,1)+(0,1,0)=(1,0,1), and $\mathcal{C}'\bullet\mathcal{P}=\mathcal{P}\bullet\mathcal{C}'=\mathcal{A}'$ corresponds to (0,1,1)+(1,1,0)=(1,1,0)+(0,1,1)=(1,0,1). Thus, [$\mathcal{D}$, $\mathcal{C}$, $\mathcal{P}$, $\mathcal{A}$, $\mathcal{D}'$, $\mathcal{C}'$, $\mathcal{P}'$, $\mathcal{A}'$] with dot products forms $Z_2 \times Z_2 \times Z_2$ group with Abelian additive operation.

We emphasize that instead of (**R**,**I**,**T**), we can consider (**I**,**M**,**T**) or (**R**,**M**,**T**) because of the relationship of **I**=**M**⊗**R**. In the case of (**I**,**M**,**T**), we have:
$\mathcal{D}$: no broken symmetry = (**I**,**M**,**T**) =(0,0,0)
$\mathcal{A}$: broken {**R**,**M**} = (**I**,**M**,**T**) =(0,1,0)
$\mathcal{C}$: broken {**I**,**M**} = (**I**,**M**,**T**) =(1,1,0)
$\mathcal{P}$: broken {**R**,**I**} = (**I**,**M**,**T**) =(1,0,0)
$\mathcal{D}'$: broken {**T**} = (**I**,**M**,**T**) =(0,0,1)

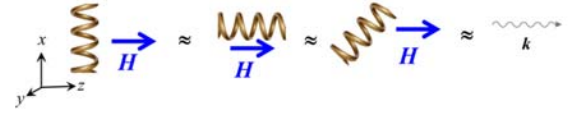

**Figure 2.** SOS relationships of a chiral spring with different orientations, magnetic field and linear momentum. In the above examples, $\mathcal{C}$ is shown with a chiral spring, magnetic field $H$ represents $\mathcal{A}'$, velocity or wave vector $k$ represents $\mathcal{P}'$, and ≈ means SOS. Note that the chiral spring has $C_\infty$ along the spring axis, which is the 1D direction; however, it can still orient in any direction for the dot product.

$\mathcal{A}'$: broken {**R**,**M**,**T**} = (**I**,**M**,**T**) =(0,1,1)
$\mathcal{C}'$: broken {**I**,**M**,**T**} = (**I**,**M**,**T**) =(1,1,1)
$\mathcal{P}'$: broken {**R**,**I**,**T**} = (**I**,**M**,**T**) =(1,0,1)
We can still have, for example, $\mathcal{C}\bullet\mathcal{P}'=\mathcal{P}'\bullet\mathcal{C}=\mathcal{A}'$= (1,1,0)+(1,0,1)=(1,0,1)+(1,1,0)=(0,1,1)= $\mathcal{A}'$. Thus, we still have $Z_2 \times Z_2 \times Z_2$ group with Abelian additive operation.
And in the case of (**R**,**M**,**T**), we have:
$\mathcal{D}$: no broken symmetry = (**R**,**M**,**T**) =(0,0,0)
$\mathcal{A}$: broken {**R**,**M**} = (**R**,**M**,**T**) =(1,1,0)
$\mathcal{C}$: broken {**I**,**M**} = (**R**,**M**,**T**) =(0,1,0)
$\mathcal{P}$: broken {**R**,**I**} = (**R**,**M**,**T**) =(1,0,0)
$\mathcal{D}'$: broken {**T**} = (**R**,**M**,**T**) =(0,0,1)
$\mathcal{A}'$: broken {**R**,**M**,**T**} = (**R**,**M**,**T**) =(1,1,1)
$\mathcal{C}'$: broken {**I**,**M**,**T**} = (**R**,**M**,**T**) =(0,1,1)
$\mathcal{P}'$: broken {**R**,**I**,**T**} = (**R**,**M**,**T**) =(1,0,1)
We can still have, for example, $\mathcal{C}\bullet\mathcal{P}'=\mathcal{P}'\bullet\mathcal{C}=\mathcal{A}'$= (0,1,0)+(1,0,1)=(1,0,1)+(0,1,0)=(1,1,1)=$\mathcal{A}'$. Thus, we still have $Z_2 \times Z_2 \times Z_2$ group with Abelian additive operation.

### 2.3 Cross products of 1D objects

A cross product (**X**x**Y**=**Z**) means that two 1D objects (**X** and **Y**) coexist in perpendicular spatial directions, and coupled to each other in such a way that they act like one (**Z**) of eight 1D objects along the 3rd perpendicular spatial direction. The full list of cross products between two vectors/directors is shown in Table 3. For example, we have $\mathcal{P}\text{x}\mathcal{P}'=\mathcal{A}'$. The cross products between vector-like 1D objects are illustrated in Fig. 3. Cross products between directors and vectors, i.e. the ones with [?] marks, do not behave like 1D objects along the 3rd perpendicular direction. For example, $\mathcal{D}\text{x}\mathcal{P}$ behaves differently under **R** with the rotational axis along the $\mathcal{D}$ direction and that along the $\mathcal{P}$ direction, so $\mathcal{D}\text{x}\mathcal{P}$ does not behave like a 1D object along the 3rd perpendicular direction.

We emphasize the sign uncertainty. For example, we can have either 𝓟x𝓪'=+𝓟' or 𝓟x𝓪'=-𝓟'; however, if we have 𝓟x𝓪'=-𝓟' as shown in the cross product Table 3, then we have 𝓪'x𝓟=+𝓟'. In addition, for example, we can have 𝓟x𝓟=+𝓪 or 𝓟x𝓟=-𝓪. In other words, symmetry consideration alone cannot fix those signs – in general, symmetry cannot fix the sign of 1D objects. However, a microscopic coupling mechanism such as spin-orbit coupling can fix, for example, the sign in 𝓟x𝓪'=+𝓟' vs. 𝓟x𝓪'=-𝓟'. Now, if we ignore +- signs, then **XxY=YxX** (i.e. commutative), and cross products are associative: **(XxY)xZ=Xx(YxZ)**. For example, we have 𝓟x𝓟'=𝓟'x𝓟=𝓪' and (𝓟x𝓟')x𝓪'=𝓟x(𝓟'x𝓪')=𝓪 if we ignore +- signs. Also note that 𝓓 acts like a unity on [𝓓,𝓒,𝓓',𝓒'] and 𝓪 acts like a unity on [𝓟,𝓒,𝓟',𝓒']. However, dot and cross products are not distributive, i.e. **Xx(Y•Z)≠(XxY)•(XxZ)** and **(XxY)•Z≠(X•Z)x(Y•Z)**.

**Table 3**, Cross products between two 1D objects when the signs of cross products are ignored

|    | 𝓓  | 𝓒  | 𝓓' | 𝓒' | 𝓪   | 𝓟   | 𝓪'  | 𝓟'  |
|----|----|----|----|----|-----|-----|-----|-----|
| 𝓓  | 𝓓  | 𝓒  | 𝓓' | 𝓒' | ?   | ?   | ?   | ?   |
| 𝓒  | 𝓒  | 𝓒  | 𝓒' | 𝓒' | ?   | ?   | ?   | ?   |
| 𝓓' | 𝓓' | 𝓒' | 𝓓  | 𝓒  | ?   | ?   | ?   | ?   |
| 𝓒' | 𝓒' | 𝓒' | 𝓒  | 𝓒  | ?   | ?   | ?   | ?   |
| 𝓪  | ?  | ?  | ?  | ?  | ±𝓪  | -𝓟  | +𝓪' | +𝓟' |
| 𝓟  | ?  | ?  | ?  | ?  | +𝓟  | ±𝓪  | +𝓟' | +𝓪' |
| 𝓪' | ?  | ?  | ?  | ?  | -𝓪' | -𝓟' | ±𝓪  | -𝓟  |
| 𝓟' | ?  | ?  | ?  | ?  | -𝓟' | -𝓪' | +𝓟  | ±𝓪  |

Now, in **XxY** with vector-like **X** and **Y**, **R** along the 3rd direction is always broken. In other words, **XxY** with vector-like **X** and **Y** always acts like a vector. Thus, we do not have to consider **R** for **XxY** with vector-like **X** and **Y**. In addition, if we ignore the +- signs for cross products, then we have:

𝓪: broken {**M**}=(**I**,**T**)=(0,0)
𝓟: broken {**I**}=(**I**,**T**)=(1,0)
𝓪': broken {**M**,**T**}=(**I**,**T**)=(0,1)
𝓟': broken {**I**,**T**}=(**I**,**T**)=(1,1)

𝓟x𝓟'=𝓟'x𝓟=𝓪' corresponds to (1,0)+(1,1)=(1,1)+(1,0)=(0,1), and 𝓪'x𝓟'=𝓟'x𝓪'=𝓟 corresponds to (0,1)+(1,1)=(1,1)+(0,1)=(1,0). Thus, [𝓪,𝓟,𝓪',𝓟'] with cross products forms $Z_2xZ_2$ with Abelian additive operation. Emphasize that this $Z_2xZ_2$ with Abelian additive operation works for the cross products since both **I** and **T** have no associated direction. Now, we choose the cross-product signs in such a way that the following cross products can be obeyed in a systematic manner as shown in Table 4. Note that we have **XxY=-YxX** (i.e. anti-commutative), and cross products are associative: **(XxY)xZ= Xx(YxZ)**. For example, we have 𝓟x𝓟'=-𝓟'x𝓟=𝓪' and (𝓟x𝓟')x𝓪'=𝓟x(𝓟'x𝓪')=𝓪. Also, note that 𝓪 acts like a unity on [𝓟, 𝓒, 𝓟', 𝓒']. For cross products, we first note that for example, 𝓟x𝓟=-𝓪 can be considered as a standard cross product. Electric field (*E*) is one example of 𝓟, and $E_xxE_y=-E_yxE_x$ and $E_xxE_y$ actually behaves like 𝓪, i.e. an axial vector, along the 3rd perpendicular direction. Magnetic field (*H*) belongs to 𝓪', so $H_xxH_y=-H_yxH_x$ and $H_xxH_y$ also behaves like 𝓪 along the 3rd perpendicular direction. In any case, we choose the signs of $E_xxE_y=-𝓪$ and $H_xxH_y=-𝓪$. In addition, $E_xxH_y=-H_yxE_x$ and $E_xxH_y$ also behaves like 𝓟' along the 3rd perpendicular direction, and we have $E_xxH_y=+𝓟'$.

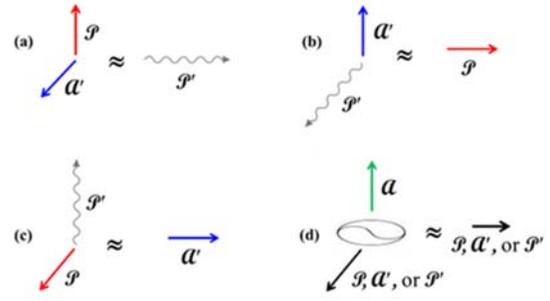

**Figure 3.** (a-c) Cross products of vector-like 1D objects among 𝓟, 𝓪', and 𝓟'. Symbols "≈" means SOS. These SOS relationships are permutable. (d) Cross products of ferro-rotation 𝓪 with 𝓟, 𝓪', and 𝓟'.

The cross-product table (Table 4) turns out to be identical with the Cayley table for $Q_8$ (Table 5), which is a non-abelian group of order eight, isomorphic to the eight-element subset [**1,i,j,k,-1,-i,-j,-k**] of the quaternions under multiplication. In the cross-product Table 4, 𝓪, 𝓟, 𝓪', and 𝓟' can correspond to **1**, **i**, **j**, and **k**, respectively. For example, 𝓟x𝓪'=-𝓪'x𝓟=𝓟' is same with **i** x **j**=-**j** x **i**=**k**. Emphasize that when we artificially choose the sign of 𝓟x𝓟=-𝓪, 𝓪'x 𝓪'=-𝓪, and 𝓟'x𝓟'=-𝓪, then the cross products follow the algebra of $Q_8$. Note that the characteristics of $Q_8$ include $(-1)^2=1$, $i^2=j^2=k^2=ijk=-1$. Correspondingly, we have -𝓪x-𝓪=𝓪, 𝓟x𝓟=𝓪'x𝓪'=𝓟'x𝓟'=(𝓟x𝓪')x𝓟'=𝓟x(𝓪'x𝓟')=-𝓪. Note also that if we have 𝓟x𝓪'=+𝓟', rather than 𝓟x𝓪'=-𝓟' then we can simply switch the role of vectors in such a way that 𝓪, 𝓪', 𝓟, and 𝓟' correspond to **1**, **i**, **j**, and **k**, respectively (see Table 6).

**Table 4**, Cross products between two 1D objects when a certain sign convention of cross products is assumed

|  | $a$ | $-a$ | $a'$ | $-a'$ | $\mathcal{P}$ | $-\mathcal{P}$ | $\mathcal{P}'$ | $-\mathcal{P}'$ |
|---|---|---|---|---|---|---|---|---|
| $a$ | $a$ | $-a$ | $a'$ | $-a'$ | $\mathcal{P}$ | $-\mathcal{P}$ | $\mathcal{P}'$ | $-\mathcal{P}'$ |
| $-a$ | $-a$ | $a$ | $-a'$ | $a'$ | $-\mathcal{P}$ | $\mathcal{P}$ | $-\mathcal{P}'$ | $\mathcal{P}'$ |
| $a'$ | $a'$ | $-a'$ | $-a$ | $a$ | $\mathcal{P}'$ | $-\mathcal{P}'$ | $-\mathcal{P}$ | $\mathcal{P}$ |
| $-a'$ | $-a'$ | $a'$ | $a$ | $-a$ | $-\mathcal{P}'$ | $\mathcal{P}'$ | $\mathcal{P}$ | $-\mathcal{P}$ |
| $\mathcal{P}$ | $\mathcal{P}$ | $-\mathcal{P}$ | $-\mathcal{P}'$ | $\mathcal{P}'$ | $-a$ | $a$ | $a'$ | $-a'$ |
| $-\mathcal{P}$ | $-\mathcal{P}$ | $\mathcal{P}$ | $\mathcal{P}'$ | $-\mathcal{P}'$ | $a$ | $-a$ | $-a'$ | $a'$ |
| $\mathcal{P}'$ | $\mathcal{P}'$ | $-\mathcal{P}'$ | $\mathcal{P}$ | $-\mathcal{P}$ | $-a'$ | $a'$ | $-a$ | $a$ |
| $-\mathcal{P}'$ | $-\mathcal{P}'$ | $\mathcal{P}'$ | $-\mathcal{P}$ | $\mathcal{P}$ | $a'$ | $-a'$ | $a$ | $-a$ |

**Table 5**, Cayley table for $Q_8$

|  | 1 | -1 | i | -i | j | -j | k | -k |
|---|---|---|---|---|---|---|---|---|
| 1 | 1 | -1 | i | -i | j | -j | k | -k |
| -1 | -1 | 1 | -i | i | -j | j | -k | k |
| i | i | -i | -1 | 1 | k | -k | -j | j |
| -i | -i | i | 1 | -1 | -k | k | j | -j |
| j | j | -j | -k | k | -1 | 1 | i | -i |
| -j | -j | j | k | -k | 1 | -1 | -i | i |
| k | k | -k | j | -j | -i | i | -1 | 1 |
| -k | -k | k | -j | j | i | -i | 1 | -1 |

## 3. Linking 1D objects and their dot/cross products to observable physical phenomena in 3D materials

The eight 1D objects and their dot or cross products are associated with particular physical phenomena such as ferromagnetism, ferroelectricity, linear magnetoelectric effects, MOKE, Faraday effect, directional non-reciprocity, etc. We, first, discuss a few examples of 1D objects exhibiting diagonal or off diagonal linear magnetoelectricity. All spin configurations in Fig. 4 are 1D objects: the spin configuration in Fig. 4(a) & (b) is an 1D object along any of the horizontal, vertical and out-of-plane directions and the 1D directions of Fig. 4(c) – (e) are along the purple arrows in Fig. 4(c) – (e). The spin configuration in Fig. 4(a) & (b) corresponds to $\mathcal{C}'$, and those in Fig. 4(c) – (e) correspond to $\mathcal{P}'$, so the spin configuration in Fig. 4(a) & (b) can exhibit diagonal linear magnetoelectricty, and those in 4(c) – (e) show off-diagonal linear magnetoelectrcity [10]. Emphasize that diagonal linear magnetoelectricty of $\mathcal{C}'$ shown in Figs. 4(a-b) is due to $\mathcal{C}'\bullet\mathcal{P}=a'$ and $\mathcal{C}'\bullet a'=\mathcal{P}$ and off-diagonal linear magnetoelectrcity of $\mathcal{P}'$ shown in Figs. 4(c-e) is due to $\mathcal{P}'\mathrm{x}\mathcal{P}=a'$ and $\mathcal{P}'\mathrm{x}a'=\mathcal{P}$. In fact, the so-called Dzyaloshinskii–Moriya interaction [18-19] can be simply represented by Fig. 4(b), and all spin configurations in Fig. 4(c) – (e) can be understood as variations of the simple magnetic toroidal moment with rotating spins in Fig. 1(k), which does exhibit off-diagonal linear magnetoelectrcity. These considerations already show the power and also the succinct nature of our 1D symmetry approach to understand and also predict emergent phenomena.

**Table 6**, Cross products between two 1D objects with the sign convention, opposite to the one in Table 4

|  | $a$ | $-a$ | $\mathcal{P}$ | $-\mathcal{P}$ | $a'$ | $-a'$ | $\mathcal{P}'$ | $-\mathcal{P}'$ |
|---|---|---|---|---|---|---|---|---|
| $a$ | $a$ | $-a$ | $\mathcal{P}$ | $-\mathcal{P}$ | $a'$ | $-a'$ | $\mathcal{P}'$ | $-\mathcal{P}'$ |
| $-a$ | $-a$ | $a$ | $-\mathcal{P}$ | $\mathcal{P}$ | $-a'$ | $a'$ | $-\mathcal{P}'$ | $\mathcal{P}'$ |
| $\mathcal{P}$ | $\mathcal{P}$ | $-\mathcal{P}$ | $-a$ | $a$ | $\mathcal{P}'$ | $-\mathcal{P}'$ | $-a'$ | $a'$ |
| $-\mathcal{P}$ | $-\mathcal{P}$ | $\mathcal{P}$ | $a$ | $-a$ | $-\mathcal{P}'$ | $\mathcal{P}'$ | $a'$ | $-a'$ |
| $a'$ | $a'$ | $-a'$ | $-\mathcal{P}'$ | $\mathcal{P}'$ | $-a$ | $a$ | $\mathcal{P}$ | $-\mathcal{P}$ |
| $-a'$ | $-a'$ | $a'$ | $\mathcal{P}'$ | $-\mathcal{P}'$ | $a$ | $-a$ | $-\mathcal{P}$ | $\mathcal{P}$ |
| $\mathcal{P}'$ | $\mathcal{P}'$ | $-\mathcal{P}'$ | $a'$ | $-a'$ | $-\mathcal{P}$ | $\mathcal{P}$ | $-a$ | $a$ |
| $-\mathcal{P}'$ | $-\mathcal{P}'$ | $\mathcal{P}'$ | $-a'$ | $a'$ | $\mathcal{P}$ | $-\mathcal{P}$ | $a$ | $-a$ |

Real 3D specimens, having SOS with those 1D objects associated with particular physical phenomena, can exhibit the physical phenomena. Emphasize that when a 3D specimen has SOS with an 1D object, the 3D specimen can have identical or lower (but not higher) symmetries than the 1D object. Linear magnetoelectricity along $z$ in $Cr_2O_3$, which we discussed earlier, is one example, and further examples of our approach using the tables of dot and cross products, practically handy for the discovery of new phenomena in low-symmetry materials, are discussed in the following.

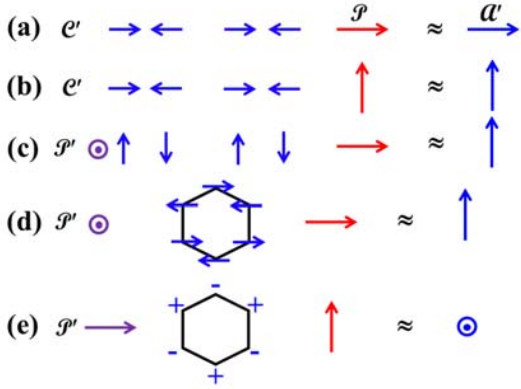

**Figure 4.** 1D objects exhibiting diagonal or off-diagonal linear magnetoelectrcity. (Red arrows depict external electric field, blue arrows are for spins or induced magnetization, and purple arrows show the orientation of $\mathcal{P}'$.) (a) & (b) linear magnetoelectricity of an antiferromagnetic dimerized spin chain with spins along the chain direction, (c) linear magnetoelectricity of an antiferromagnetic dimerized spin chain with spins perpendicular to the chain direction, (d) & (e) linear magnetoelectrcity of two types of antiferromangetic spins on a honeycomb lattice. Spin configurations of (a) & (b) correspond to $\mathcal{C}'$, and those of (c) – (e) correspond to $\mathcal{P}'$.

(1) Spin configurations in Fig. 5(a)–(c) represent helical spin state (proper screw-type spiral spin state), "toroidal moment + a canted moment", and "magnetic quadrupole moment + alternating canted moments", respectively, and all have SOS with $\mathcal{C}$. Helical spin states have been observed in numerous magnets such as centrosymmetric compounds such as YMn$_6$Sn$_6$[20] Gd$_2$PdSi$_3$[21], MnP[22], and some rare earth metals [23] and crystallographically-chiral compounds such as Co$_7$Zn$_7$Mn$_6$[24], MnSi[25], MnGe[26], Fe$_{1-x}$Co$_x$Si[27] with B20 structure, Cu$_2$OSeO$_3$[28], Cr$_{1/3}$TaS$_2$[29], etc. "toroidal moment + a canted moment" has been obsrved in crystallograpically-chiral BaCoSiO$_4$ [30], and "magnetic quadrupole moment + alternating canted moments", has been reported in crystallograpically-chiral Pb(TiO)Cu$_4$(PO$_4$)$_4$[31] and Er$_2$Ge$_2$O$_7$[32]. Note that both helical spin state and "toroidal moment + a canted moment" are 1D objects, and belong to $\mathcal{C}$,, but "magnetic quadrupole moment + alternating canted moments" is not 1D objects since its behavior under $C_2(x)$ is different from that under $C_2(xy)$; however, it does have SOS with $\mathcal{C}$.

It turns out that circularly polarized light and vortex light beams with orbital angular momentum are, in fact, chiral (i.e. all mirror symmetries with a combination of any spatial rotation are broken), so can couple with those chiral spin textures or states as well as crystallographic chirality. Thus, those chiral spin textures or states can exhibit magnetic version of natural optical activity, circular dichroism, etc. This was first proposed in Ref. 9, and in fact, a natural optical activity in a THz range has recently been observed in the helical spin state of CuO with an achiral crystallographic structure [33]. Similar optical activities of "toroidal moment + a canted moment", and "magnetic quadrupole moment + alternating canted moments" have never been studied, and can be confirmed experimentally in the future. In addition, vortex light beams should couple with either crystallographic chirality or chiral spin textures/states, but these effects have never been studied, and can be investigated in the future. Both crystallographic chirality and magnetic chirality can exhibit magnetochiral effects, i.e., directional nonreciprocal effects along the magnetic-field directions in the presence of external magnetic fields. Directional nonreciprocal spin waves in magnets with helical spins (e.g. Er metal with centrosymmetric crystallographic lattice) have been studied [34], but study of directional nonreciprocal spin waves in "toroidal moment + a canted moment" or "magnetic quadrupole moment + alternating canted moments" has never been reported, and can be a future topic.

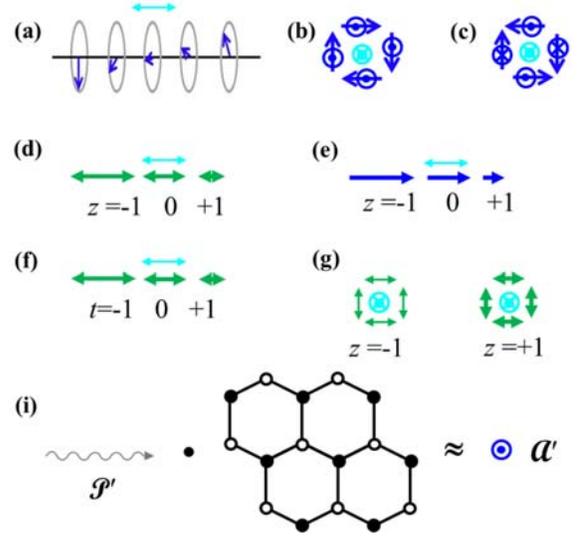

**Figure 5.** Various 1D objects: (a) helical spin state, (b) magnetic toroidal moment + a canted moment, (c) magnetic quadrupole moment + alternating canted moments. Blue arrows: spins or magnetic fields, green double arrows: uniaxial strains, light-blue double arrows: relevant 1D directions (i.e. the $z$ axes). (d) Spatial variation of strain, (e) spatial variation of magnetic field, (f) temporal variation of strain, and (g) temporal variation of strain with 1D direction along the out-of-plane direction. "$t$" is time and "$z$" is the $z$ coordinate. (i) $\mathcal{P}'$ combined with the non-centrosymmetric Fig. 1(m) state, which is not an 1D object, can result in magnetization or $\mathcal{a}'$.

(2) Spatial variation of strain in Fig. 5(d) and (g), temporal variation of strain in Fig. 5(f), and spatial variation of magnetic field in Fig. 5(e) belongs to $\mathcal{P}$, $\mathcal{A}'$, and $\mathcal{C}'$, respectively. Since we have $\mathcal{P} \cdot \mathcal{C}' = \mathcal{A}'$, and $\mathcal{A}' \cdot \mathcal{C}' = \mathcal{P}$, we can have these:

(a) Spatial variation of strain shown in Fig. 5(d) or Fig. 5(g) on linear magnetoelectric $Cr_2O_3$ can produce a non-zero net magnetic moment. These phenomena correspond to flexomagnetism. In fact, this effect should work in any linear magnetoelectrics, so can be tested in numerous linear magnetoelectrics such as $Mn_4Nb_2O_9$[35], $LiCoPO_4$[36], $TbPO_4$[37] etc. This effect by strain gradient should work for both diagonal linear magnetoelectrics as well as off-diagonal linear magnetoelectrics even though the induced magnetization direction with respect to the strain gradient direction should be different. There exist a number of compounds forming in $\bar{1}'$ such as $BaNi_2(PO_4)_2$ [38], $MnPSe_3$[39], $CaMnGe_2O_6$[40], $NaCrSi_2O_6$ [41], $YbMn_2Sb_2$ [42], $CaMn_2Sb_2$ [43], and these compounds do exhibit diagonal linear magnetoelectrics as well as off-diagonal linear magnetoelectrics, so strain gradient experiment on these compounds can be an interesting future research topic.

(b) Spatial variation of magnetic field shown in Fig. 5(e) belongs to $\mathcal{C}'$, so it combined with magnetic field can produce an electric polarization or a voltage gradient. It combined with electric field or polarization can result in magnetization. These effects should, in principle, work in any materials, but have never been reported. One way to understand the mechanism behind voltage gradient induced by magnetic field gradient combined with magnetic field is: electrons are magnetized due to magnetic field, and magnetic field gradient exerts force on those magnetized electrons, so electron density gradient, i.e. voltage gradient, is induced.

(c) Temporal variation of strain in Fig. 5(f) on any materials can produce a non-zero net magnetic moment remarkably. This effect should work in any materials, magnitude of induced magnetization is expected to be proportional to the magnitude of temporal strain gradient, and large temporal strain gradient can be readily induced by, for example, fast optical pumping.

(d) Temporal variation of strain in Fig. 5(f) on any linear magnetoelectrics such as $Cr_2O_3$, $Mn_4Nb_2O_9$, $LiCoPO_4$, $TbPO_4$, etc. can lead to a non-zero net electric polarization.

(e) We also note that spatial thermal gradient is just like spatial strain gradient in terms of symmetry, and belongs to $\mathcal{P}$, which is, In fact, the origin of Seebeck (thermoelectric power) effect. Therefore, spatial thermal gradient can do what spatial strain gradient can do. For example, thermal gradient on any linear magnetoelectrics such as $Cr_2O_3$, $Mn_4Nb_2O_9$, $LiCoPO_4$, $TbPO_4$, etc. should results in magnetization. This effect has never been reported and can be coined as thermomagnetic power effect.

(3) Cycloidal spin state is $\mathcal{P}$ in the direction perpendicular to the spin modulation direction, but in the spin plane, so can induce polarization (voltage gradient), which has been observed in frustrated magnets such as $TbMnO_3$ [44], $LiCu_2O_2$ [45], $LiCuVO_4$ [46] etc. Existing polarization in, for example, $BiFeO_3$ can induce cycloidal spin order as an inverse effect [47]. It is experimentally shown that the spin rotation direction (clockwise vs. counter-clockwise) of cycloidal spins can be controlled by external electric fields. Now, a spatial thermal gradient has the same symmetry with polarization, so it is possible that cycloidal spins can induce a thermal gradient and also a thermal gradient can control the spin rotation direction of cycloidal spins, which need experimental confirmation in the future.

(4) $\mathcal{A}' \times \mathcal{P}' = \mathcal{P}$ means that the simultaneous presence of current and magnetic field along a direction perpendicular to the current direction acts like electric field or voltage gradient along the 3$^{rd}$ perpendicular direction. This can be well utilized for various novel experiments – two examples are listed below: It was reported that cycloidal spin order comes in in centrosymmetric metallic $CrB_2$ [48] at low temperatures – cycloidal spins are on the [110]-[001] plane and the modulation direction is along [110]. Since $CrB_2$ is non-polar, cycloidal spin states can be either clockwise- or counterclockwise-types; however, the current along [110] and magnetic field along [1$\bar{1}$0] can favor one of clockwise- or counterclockwise-types, and switching one of current and magnetic field can favor the other type. Applying magnetic field on the helical spin state of centrosymmetric $YMn_6Sn_6$ [22] was reported to lead to a transverse conical spin state, which is a combination of net moment state with magnetization along the magnetic field direction, and cycloidial spins state with modulation along *c* (cycloidal spins are perpendicular to the magnetic field direction). Then, current along *c* in a combination of magnetic field along [1$\bar{1}$0] can favor one of clockwise- or counterclockwise-types of cycloidal spin state, and flipping the current direction can favor the other type.

(5) Since we have, for example, $\mathcal{P} \times \mathcal{A} = \mathcal{P}$, applying external electric field ($\mathcal{P}$) on ferro-rotation ($\mathcal{A}$) can induce a tilting of the total electric field or a transverse (Hall-type) voltage (𝓟) without any external magnetic field. Similarly, we have 𝓪′x𝓪=𝓪′, so applying external magnetic field (𝓪′) on a ferro-rotational (𝓪) systems can induce a tilting of the total magnetic field or a transverse magnetic field (𝓪′). These effects can be tested in numerous ferro-rotational systems such as Ni(Fe)TiO$_3$ [49], RbFe(MoO$_4$)$_2$ [50], CaMn$_7$O$_{12}$ [51], Cu$_3$Nb$_2$O$_8$ [52], NbO$_2$ [53], and Pb$_2$CoTeO$_6$ [54], 1T-TaS$_2$ with commensurate charge-density wave [55], etc.

(6) The experimental setup for optical activities, including the typical Faraday rotation of linearly-polarized light transmission, natural optical activity, circular dichroism, vortex beam dichroism, and even circular-polarized photogalvanic effect (CPGE) has broken {**M**,**I**⊗**T**} [9]. For 1D objects, having unbroken **R**, **M**=**I**⊗**R**; thus, broken {**I**⊗**R**,**I**⊗**T**} is the requirement for optical activities. Thus, among eight kinds of 1D objects, chirality (𝓒=(0,1,0)) or axiality′ (𝓪′=(1,0,1)) objects do have broken {**I**⊗**R**,**I**⊗**T**}, so can exhibit optical activities. As discussed earlier, all chiral spin textures/states should exhibit magnetism-induced natural optical activity, and vortex beams should couple with crystallographic chirality as well as chiral spin textures/states. The requirement for MOKE along $z$ is broken {**C$_3$(x)**,**C$_3$(y)**,**M$_x$**,**M$_y$**,**T**}[9]. As discussed above, this requirement becomes broken {**I**⊗**R**,**T**} for MOKE along 1D directions of 1D objects. Thus, among eight kinds of 1D objects, any chirality′ (linear magnetoelectrics, 𝓒′=(0,1,1)) or axiality′ (𝓪′=(1,0,1)) objects can exhibit MOKE. Any axiality′ (𝓪′) objects, basically showing non-zero net magnetic moments, can exhibit both optical activities effects and MOKE. Any diagonal linear magnetoelectrics such as Cr$_2$O$_3$, Mn$_4$Nb$_2$O$_9$, and TbPO$_4$, should exhibit MOKE.

(7) 𝓟x𝓟′=𝓪′ corresponds to the very first experiment of a magnetoelectric effect in a solid in 1888 by Wilhelm Röntgen [56], who showed that a dielectric material moving through an electric field would become magnetized. It also means that electric current flow along a direction perpendicular to polarization in a polar system can induce magnetization along the 3$^{rd}$ perpendicular direction, and this magnetization with current can induce a Hall effect along the polarizatin direction without external magnetic field. This Hall effect is quadratic to current magnitude since the induce magnetization as well as current itself are proportional to the current. These effects have been recently observed in ultrathin WTe$_2$ and strained monolayer MoS$_2$ [57-59]. It will be imperative to observe the relevant effects in bulk polar conductors such as GeTe [60] and Ca$_3$Ru$_2$O$_7$ [61] in the future.

(8) 𝓟′•𝓪′=𝓒 means that the simultaneous presence of electric current (𝓟′) and magnetic field (𝓪′) along one direction does have chirality (𝓒), so can couple of crystallographic chirality or chiral spin textures/states. Centrosymmetric systems such as YMn$_6$Sn$_6$ [20], Gd$_2$PdSi$_3$ [21], MnP [22], and some rare earth metals can show helical spin states, which are chiral. If electric current and magnetic fields are applied to one direction, then it can favor one chirality vs. the other chirality of helical spins, and reversing one of current and magnetic field directions will results in the flipping of chirality of helical spins [22].

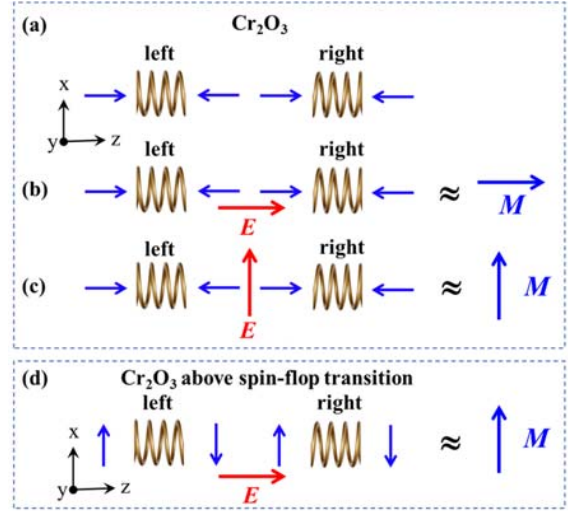

**Figure 6.** (a) A schematic of the magnetic state of Cr$_2$O$_3$, having SOS with 𝓒′ along any of $x$, $y$ and $z$. Left and right springs represent the crystallographic configurations of oxygen environments [10]. (b) The induced surface magnetization $M_z$ (blue arrow) in the presence of electric field $E_z$ (red arrow) through 𝓒′•𝓟= 𝓪′. (c) The induced surface magnetization $M_x$ (blue arrow) in the presence of electric field $E_x$ (bold red arrow) through 𝓒′•𝓟= 𝓪′. (d) A schematic representing the magnetic state of Cr$_2$O$_3$ above spin flop transition in $H_z$, and having SOS with 𝓟′ along $y$. In this state, magnetization $M_x$ is induced by electric field $E_z$ through 𝓟′x𝓟=𝓪′.

(9) Emphasize that one can readily come up with numerous other phenomena in real materials using the tables of dot (Table 2) and cross products (Table 3) of 1D objects, which have rarely been observed or even discussed. We also note that our dot and cross product approach is applicable for non-1D objects. For example, 𝓟′ combined with the non-centrosymmetric Fig. 1(m) state, which is not an 1D object, does have SOS with 𝓪′, so can result in magnetization (Fig. 5(i)). Complex phonon behavior reported recently in Ref. [62] can be simply understood with this SOS relationship. In addition, 𝓟′ combined with 𝓪′ can lead to 𝓟 through 𝓟′x𝓪′=𝓟, which is perpendicular to 𝓟′ and 𝓪′. Furthermore, the SOS relationship shows that -𝓟′ leads to -𝓪′ and +𝓟; thus, this effect is non-linear, for example, the induce $\mathcal{P}$ goes like $(\mathcal{P}')^2$. This behavior not been observed and is a topic for future experimental confirmation.

(10) Finally, since $Cr_2O_3$ is relevant to many phenomena that we have discussed, let's summarize what can be observed in $Cr_2O_3$. The magnetic space group is $R\bar{3}'c'$, which has unbroken {$R_x$,$C_3(z)$,$I\otimes T$,$C_3(z)\otimes I\otimes T$,$M_x\otimes T$} and is non-1D objects along $x$, $y$ or $z$. Fig. 6(a) illustrates the left-chiral and right-chiral arrangements of the crystallographic oxygen environments and the corresponding spin states of $Cr_2O_3$. As we discussed earlier, it has broken {$I$,$T$,$M_x$,$M_y$,$R_x\otimes T$,$R_y\otimes T$}, so has SOS with $\mathcal{C}'$ along $z$, so exhibit linear magnetoelectricity along $z$ (relevant to $\mathcal{C}'\bullet\mathcal{P}=\mathcal{A}'$ and $\mathcal{C}'\bullet\mathcal{A}'=\mathcal{P}$). In fact, it has also broken {$I$,$T$,$M_y$,$M_z$,$R_y\otimes T$,$R_z\otimes T$}, and {$I$,$T$,$M_x$,$M_z$,$R_x\otimes T$,$R_z\otimes T$}, so does also exhibit linear magnetoelectricity along $x$ and $y$. Since we have broken {$M_x$,$M_y$,$T$} in $Cr_2O_3$, we can have MOKE in $Cr_2O_3$ along $z$ [63]. We also note that surfaces of $Cr_2O_3$ can have an effective electric field, i.e. $\mathcal{P}$, and we have $\mathcal{C}'\bullet\mathcal{P}=\mathcal{A}'$, so surfaces of $Cr_2O_3$ can have out-of-plane magnetization. Fig. 6(b) and Fig. 6(c) illustrate the induced surface magnetization in the presence of surface electric field, leading to optical activity (circular dichroism) or MOKE. On the other hand, our symmetry analyisis, $\mathcal{P}\bullet\mathcal{C}'=\mathcal{A}'$, predicts strain gradient or thermal gradient on $Cr_2O_3$ can produce magnetization along the stain gradient or thermal gradient direction – this can happen along any of $x$, $y$, and $z$ since both strain gradient and thermal gradient are $\mathcal{P}$-type. In addtion, since temporal variation of strain is $\mathcal{A}'$ and we have $\mathcal{C}'\bullet\mathcal{A}'=\mathcal{P}$, temporal strain gradient on $Cr_2O_3$ can induce polarization along the temporal strain gradient direction – again this can work along any of $x$, $y$, and $z$. Note that the surface magnetization induced by an surface effective electric field should work presumably in any linear magnetoelectrics [64]. Fig. 6(d) represents the magnetic state of $Cr_2O_3$ above spin-flop transition in $H_z$. This spin-flopped state has magnetic toroidal moment along $y$, so exhibits off-diagonal linear ME through $\mathcal{P}'\times\mathcal{P}=\mathcal{A}'$. With surface electric field $E$ as shown in Fig. 6(d), this spin state allows in-plane surface magnetization.

## 4. Summary and outlook

In summary, we, first, defined the exact meaning of 1D objects in terms of symmetry. These 1D objects include (magnetic) lattices with periodic structures, traditional vectors/directors such as electric field, magnetic field, velocity and uniaxial strain, as well as 3D objects with particular shapes but no internal structures (i.e. no periodic lattices). The 1D objects can be classified into eight different kinds, four of which are vector-like and the other four are director-like in a rotation symmetry consideration around the direction perpendicular to the 1D direction. Dot products and also cross products can be physically defined. The dot products form a $Z_2xZ_2xZ_2$ group with Abelian additive operation, and the cross products form a $Z_2xZ_2$ group with Abelian additive operation when the signs of the cross products are ignored. We have also shown that $Q_8$, which is a non-abelian group of order eight, can isomorphic to the cross products when the signs of the cross products are systematically chosen. The group theoretical consideration of 1D objects in this paper, combined with the SOS concept to link 3D materials/measurements and 1D objects, will be an important guidance in unveiling new phenomena in materials with known symmetry properties or discover new materials with desired physical phenomena. Finally, we note that our symmetry approach does not tell neither the microscopic mechanism for physical phenomena nor their magnitudes. However, it turns out that, for example, all magnets with cycloidal spin order, which is $\mathcal{P}$, always show experimentally measurable polarization. Thus, these situations appear to resonate with the famous statement of Murray Gell-Mann: "everything not forbidden is compulsory" [65]. In other words, when a certain physical phenomenon is allowed by the consideration of broken symmetries, the effect is often large enough, so it is experimentally observable. Thus, our symmetry approach is, at least, an essential guidance, and educated ingredients such as, for example, large spin-orbit coupling (with heavy elements) for off-diagonal effects, such as cycloidal spin-induced multiferroicity through antisymmetric exchange coupling, off-diagonal linear magnetoelectricity, or MOKE, always help to identify new materials for desired phenomena.

**Acknowledgements**
The work at Rutgers University was supported by the DOE under Grant No. DOE: DE-FG02-07ER46382.REFERENCES

**Data availability statememt**
All study data are included in the article.